\documentstyle[aps,preprint]{revtex} 
\tightenlines 
\begin{document}

\title{Nonuniversality in the pair contact process with diffusion} 
 
\author{Ronald Dickman$^\dagger$ and
Marcio Argollo Ferreira de Menezes$^*$
} 
\address{ 
$^\dagger$Departamento de F\'{\i}sica, ICEx, 
Universidade Federal de Minas Gerais,\\
Caixa Postal 702, 
30161-970 Belo Horizonte, MG, Brazil\\ 
$^*$Department of Physics, University of Notre Dame,
Notre Dame, IN 46556} 
\date{\today} 
 
\maketitle 
\begin{abstract} 
We study the static and dynamic behavior of the one dimensional pair contact
process with diffusion.  Several critical exponents are found
to vary with the diffusion rate, while the order-parameter moment ratio
$m=\overline{\rho^2} /\overline{\rho} ^2$ grows logarithmically with the
system size.  The anomalous behavior of $m$ is traced to a violation of scaling in the
order parameter probability density, which in turn reflects the presence of {\it two
distinct sectors}, one purely diffusive, the other reactive, within the active phase.  
Studies restricted to the reactive sector yield precise estimates
for exponents $\beta$ and $\nu_\perp$, and confirm finite size scaling of the order 
parameter.  In the course of our study we determine, 
for the first time, the universal value $m_c = 1.334$ associated with the
parity-conserving universality class in one dimension.
\end{abstract} 
 

 \newpage

The pair contact process (PCP) \cite{pcp1,pcp2} is a nonequilibrium stochastic
model which, like the basic contact process (CP) \cite{harris,liggett,marro}, 
exhibits a phase transition to an absorbing state.  While the absorbing state in
the contact process corresponds to a unique configuration (an empty lattice), the PCP possesses
infinitely many absorbing configurations.  Numerical  and theoretical
studies nevertheless indicate that the PCP belongs to the same universality class as the CP
(namely, that of directed percolation (DP)), but with anomalies in the critical spreading dynamics
\cite{pcp1,pcp2,rdjaff,jaffrd,odor98,pcpcmp,inac,gcr,munozjsp}.  An infinite number of absorbing
configurations arise in the PCP because all processes (creation and annihilation), require
a nearest-neighbor (NN) pair of particles (to be referred to simply as a ``pair" in what follows).
If individual particles are allowed to hop on the lattice, however, there are but two absorbing states: the empty lattice, and the state of a single particle hopping.

Study of the diffusive pair contact process (PCPD) was stimulated by the observation of
Howard and T\"auber \cite{howard} that its Langevin description would involve complex
noise (this in contradistinction to the CP and allied models (real noise) and the
parity-conserving class (imaginary noise)).  
On the basis of numerical 
results in their pioneering density-matrix renormalization group study,
Carlon et al. \cite{carlon}, noted that certain critical exponents in the PCPD had
values similar to those known for the   
parity conserving (PC) universality class.
Hinrichsen \cite{hh} reported simulation results 
inconsistent with the PCPD being in the parity conserving class, and instead proposed that the 
model defines a {\it distinct} class.
In particular, while models in the PC class possess two symmetric absorbing states, the two
absorbing states of the PCPD are not related by any symmetry.  Interestingly, Park et al. found
that even when such a
symmetry is imposed on the PCPD, its critical exponents remain different from those of the PC 
class \cite{park}.  The distinctive behavior of the PCPD was further confirmed in simulations
by \'Odor \cite{odor2000}, who presented evidence for the existence of two universality
classes (for diffusion probabilities greater than, or less than, about 0.3). 
Henkel and Schollw\"ock, on the other hand, suggested, on the basis of a study of universal
finite-size scaling amplitudes, that for finite diffusion rates, the critical behavior of the 
PCPD belongs to a single universality class \cite{henkel01}.
Our goal in this Rapid Communication is to shed some light on this rather confusing situation
by studying moment ratios and probability distributions in the critical PCPD.

The PCP is defined on a lattice, with each site either
occupied (by a ``particle") or vacant.  Only pairs
of occupied sites exhibit activity; each has a rate of $p$
of mutual annihilation, and a rate of $1\!-\!p$ to
create a new particle at a NN site, if this site (chosen at random) is
vacant.  
For $p > p_c$ ($\simeq 0.077\;090(5)$ in 1-$d$ \cite{rdjaff}), the system falls into the absorbing
state (all activity ceases).  The order parameter 
is the density of pairs.

In the PCPD, in addition to the creation and annihilation processes described above,
each particle attempts to hop, at rate $D$, to a randomly chosen NN site;
the move is accepted if the target site is vacant.
The model again exhibits a continuous transition to the absorbing state, at a critical
annihilation rate $p_c(D)$ that increases with the diffusion rate. 
Once particles are allowed to diffuse, the nature of the system changes radically.
The absorbing state is modified as noted above, and the
order parameter is now the
particle density not the pair density.  
In contrast to simpler models like the CP, in which diffusion does not alter the
critical behavior \cite{ud,dcpser}, diffusion represents a {\it singular perturbation}
in the pair contact process, since any $D > 0$ implies a fundamental change in the
phase structure and in the identity of the order parameter.

We perform extensive simulations of the one-dimensional PCPD, using
systems of L = 20, 40,...,1280 sites, with durations of 
10$^4$ - 4 $\times $ 10$^6$ time steps, and sample sizes of 10$^4$ -
10$^6$ realizations.  
Initially all sites are occupied.
We determine the mean particle density $\overline{\rho}$,
and pair density $\overline{\rho_p}$,
the moment ratio $m = \overline {\rho^2}/\overline{\rho}^2$,
and the survival probability
$P_s(t)$.  
(The overline denotes a stationary average.)
The exponential decay of the latter
permits us to determine the lifetime $\tau$.  We
concentrate on the critical region, $p \simeq p_c(D)$.

Experience with absorbing-state phase transitons leads us to expect the
following scaling properties at the critical point:
$\overline{\rho} \sim L^{-\beta/\nu_\perp}$;
$
\tau \sim L^{\nu_{||}/\nu_\perp}
$; 
and
$
m \to m_c
$, a universal critical value \cite{rdjaff}.  
We use power-law dependence of $\rho$
on system size to determine the critical annihilation rate $p_c (D)$.
For comparison we applied the same algorithm to the parity-conserving
branching-annihilating random walk (BAW) model studied by 
Zhong and ben-Avraham \cite{zhong}.

Fig. 1 shows the scaling of the order parameter
with system size, at the critical point, for the PCPD and the BAW;  
in the PCPD, $\beta/\nu_\perp$  decreases with increasing
diffusion rate.  (The fact that the 
data points for the PCPD with $D\!=\!0.5$ and the BAW are nearly
identical appears to be a coincidence, since the scaling of the
relaxation time $\tau$ is quite different in the two cases.)
Fig. 2 shows that while the moment ratio $m$ 
attains a limiting value in the BAW model, it
{\it grows} with $L$ in the PCPD (roughly, $\sim \ln L$), a most
unusual behavior.
We find $m_c$ = 1.3340(4) for the BAW model,
while $m_c$= 1.1735(5) for the directed percolation class in 1+1 dimensions \cite{rdjaff}.

In models with an absorbing-state phase transition, the
probability distribution for the order-parameter, $P(\rho;L)$
is expected to exhibit scaling at the critical point,
\begin{equation}
P(\rho;L) = \overline{\rho} \;{\cal P} (\rho/\overline{\rho} ),
\label{scop}
\end{equation}
where ${\cal P} $ is a normalized scaling function,
as was verified for the PCP without diffusion \cite{pcp2}.  In the present case,
the steady growth of $m_c$ with system size implies that $P(\rho;L)$ does not
obey scaling. 
The particle and pair probability distributions, shown (for $D \!=\! 0.1$)
in Fig. 3, evidently do not scale.
Instead, the most probable value of the particle number is always 2
(configurations with fewer than two particles are of course
absorbing), and the overwhelmingly most probably number of pairs is
{\it zero}, independent of system size.  The distributions exhibit a
tail that grows broader with increasing system size; these ``tail events"
are responsible for the observed critical behavior.  The tails, which
have a Gaussian form, again violate the
scaling of Eq. (\ref{scop}).  (The pair distribution exhibits a second
maximum, away from $\rho_p = 0$, whose position increases 
slowly with system size, roughly as $L^{0.6}$.)

The particle and pair probability distributions confirm lack of 
scaling, and, perhaps more importantly,
provide a clue to the enigmatic behavior of the process.  In the PCP
without diffusion, there is always at least one pair present in the active state.
But once we add diffusion, being in the active (i.e., non-absorbing) 
state implies that there are at least two particles, but not neccessarily any
pairs.  At $p_c$, the process apparently favors 
configurations with a small number of particles, but with no pairs.
(For $D\!=\!0.1$, for example, the probability of having no pairs
remains at about 0.8 for the the system sizes studied here, and shows
no sign of decreasing as $L$ grows; for $D\!=\!0.5$ this probability is
about 0.58, and for $D\!=\!0.85$, about 0.5.)
While in this ``purely diffusive" sector, the activity is that of a set of random
walkers, but the particle number does not change, and critical
fluctuations are not generated.  From time to time the system
ventures into the sector with a nonzero pair number (the ``reactive sector"), 
and may there exhibit a burst of creation and annihilation reactions.
We expect the latter activity to possess scale invariance at $p_c$.  Thus
the probability distribution may be seen as a superposition of
distributions associated with the two sectors.  
In this light, lack of scaling is quite understandable.
In the purely diffusive sector, the particle-number distribution is
highly-peaked at $n\!=\!2$, with (for $D\!=\!0.1$) a mean value
of about 3.5, independent of system size. 
(For $D\!=\!0.5$ and 0.85, the mean particle number in the purely diffusive sector is about 3.2).

These observations motivate us to {\it exclude} the purely diffusive
sector by studying properties {\it conditioned on having at least
one pair} in the system.  Note that this does not
modify the dynamics of the system in any way; we simply restrict
the averages to configurations having one or more pairs.  Fig. 4 shows the
order parameter distribution in the reactive sector, plotted
in the reduced variables $\rho* = \rho/\overline{\rho}$ and
$P^* = \overline{\rho} P$, for the same
parameter values as in Fig. 3.  The distribution now assumes a form
very similar to that found in the nondiffusive PCP \cite{pcp2},
with a maximum at a nonzero value of the order parameter,
and shows evidence of scaling.  Thus the behavior in the reactive
sector is much closer that familiar from the
contact process, the PCP, and related models with an absorbing state
phase transition.

Closer examination reveals, however, that the scaling collapse is
imperfect. Studies of larger systems confirm that the maximum of the
scaled order parameter distribution gradually shifts to smaller values of 
$\rho^*$, and that the distribution becomes broader, with increasing $L$.  (The latter
is evident in the results for $m$ discussed below.)  While we do not
claim to have a complete understanding of this ``defect," a possible
explanation is that for large $L$, configurations with but a single
pair represent a system with only a small reactive region, the
remainder residing in the purely diffusive sector.  We defer a full investigation of
this rather subtle question to future work.

Once we restrict the sample to the reactive regime, we eliminate a
large source of uncertainty (i.e., the erratic switching between the two
sectors), and are able to obtain more precise results. 
Using, as before, the criterion of power-law dependence of 
$\overline{\rho}$ on system size, we determine the critical parameter
$p_c$ and the ratio $\beta/\nu_\perp$ to good precision; these values are
given in Table I.  Restricting the averages to the reactive sector
changes the value of $p_c$ by 0.1\% or less.  There are more pronounced
changes in $\beta/\nu_\perp$: without the restriction, we obtain
0.585, 0.50, and 0.465 for $D\!=$ 0.1, 0.5 and 0.85, respectively.
(We regard these as poorer estimates, colored by the superposition
of the two sectors.  Note however that these values exhibit the same
trend - decreasing $\beta/\nu_\perp$ with increasing diffusion rate -
as observed in the reactive sector.)   
Fig. 4 (inset) shows the critical
moment ratio $m_c$ versus system size, in the reactive sector.
Its value is now comparable (for the system sizes studied here),
to that for the DP and PC classes, but a
slow growth (roughly linear in $\ln L$) is again evident.
(Restricting the sample to configurations with two pairs
leads to a reduction in $m$, but not in its rate of growth
with system size.)

A possible weak point in our analysis is that we assume finite size scaling
(i.e., the power-law dependence of $\overline{\rho}$ on
system size), in determining $p_c$, whilst the results for $m$ indicate
that there is still a (relatively weak) violation of scaling.
We therefore check our method by studying the order parameter (again
restricted to the reactive sector), in the supercritical regime, $p < p_c$.
We verify that the order parameter follows a power law,
$\overline{\rho} \sim (p_c \!-\! p)^\beta$, and in so doing obtain
the estimates for $\beta$ given in Table I.  
This exponent decreases steadily with $D$, as found in Ref. \cite{odor2000}.
(A direct comparison with the results of Ref. \cite{odor2000}
is not possible since the latter study uses a parallel-update scheme, in contrast
to the sequential updating used here.)

In fact, our results
verify finite size scaling for the order parameter, i.e., the relation,
\begin{equation}
\overline{\rho} = L^{-\beta/\nu_\perp} {\cal R} (L^{1/\nu_\perp} \Delta),
\label{rhofss}
\end{equation}
where $\Delta = p_c \!-\! p$ and the scaling function
${\cal R}(x) \sim x^\beta$ for $x \gg 1$.  The data collapse is
evident in Fig. 5.  From this analysis we obtain
$\nu_\perp = 1.10$, 1.09, and 1.10 for $D\!=\!0.1$, 0.5 and 0.85,
respectively, suggesting that this exponent does not vary with
the diffusion rate.

We also studied the decay of the particle density starting from
a fully occupied lattice at the critical point, restricting the
sample to the reactive sector.  (In the early stages of the evolution,
the probability for the system to be in the reactive sector is
nearly unity, but at later times this probability decays much more
rapidly than the survival probability itself.)  From a data-collapse
analysis of $\rho(t)$, using the finite-size scaling form,
$\rho = L^{-\beta/\nu_\perp} {\cal F} (t/L^{\nu_{||}/\nu_\perp})$,
we obtain the estimates for $z = \nu_{||}/\nu_\perp$ listed in Table I.
(The corresponding estimates, without the restriction to the reactive
sector are: 1.87(1) for $D\!=\!0.1$, 1.82(1) for $D\!=\!0.5$ and 0.85.)

We complement our analysis with a study of dynamic
properties, using a parallel-update scheme.  (Details of the method will
be reported elsewhere \cite{mard}.)  We determine the exponent
$\delta$ from the decay of the particle density, starting with all
sites occupied: $\rho \sim t^{-\delta}$.  The exponent $\eta$ is determined
from the growth in the number of active sites, starting from a single
pair: $n(t) \sim t^\eta$.  The results (based on samples of $10^4$ realizations, for 
systems of 1280 sites, without restricting the sample to the
reactive sector), shown in Table II, indicate that these
exponents also depend on the diffusion rate, and again are very different from 
those of the BAW class.  Our results for $\delta$ and $\eta$ are similar to
those obtained by \'Odor \cite{odor2000}, although a direct numerical
comparison is not possible, owing again to differences in the updating
scheme.

In summary, we have performed extensive studies of the PCPD, including
the probability distributions for the order parameter and number of pairs.
Our results clearly exclude the model from both the parity-conserving
and the DP universality classes, supporting
Hinrichsen's proposal that the model belongs to a distinct
class.  The criticial exponents $\beta$, $\eta$ and $\nu_{||}$ vary 
with the diffusion rate, while $\nu_\perp$ appears to be independent 
of this parameter.
An interesting open question is whether the PCPD can be described by a
single universality class (with unusually strong corrections to scaling yielding 
an apparent variation of critical exponents on $D$) \cite{henkel01}, two
distinct universality classes (one for high diffusion rates, the other for
low, but finite $D$), as suggested by \'Odor \cite{odor2000}, or
even exponents that vary continuously with $D$.
Our data are not sufficient to distinguish between these hypotheses.
We note, however, that
we observe relatively little change in the exponent values for 
$D\!=\!0.5$ and 0.85, compared with the changes between $D\!=\!0.1$ and 0.5.
A similar observation applies to the size dependence of $m$ shown in Fig. 4.

The growth of the moment ratio $m$ with system size signals
a violation of scaling in the associated probability distribution, 
which we have argued is a consequence
of there being two sectors, one reactive, the other purely diffusive, 
within the active phase.  Restricting averages to the
reactive sector, we find good evidence of finite size scaling of the
order parameter, and a much weaker violation of scaling for the
probability distribution.  
The question of how this remaining violation may be eliminated is
an important subject for future investigation.
We expect that
decomposition of configuration space into sectors will prove
useful in understanding other systems exhibiting bursts of activity
separated by long quiescent periods.

\vspace{2em}

\noindent{\bf Acknowledgments} 
\vspace{1em} 

We are grateful to Miguel A. Mu\~noz, Geza \'Odor, and Malte Henkel for 
valuable comments and suggestions.  
This work was supported by CNPq, and CAPES, Brazil. 
\vspace{2em}

\noindent$^\dagger$ e-mail: dickman@fisica.ufmg.br \\
\noindent$^*$ e-mail: mdemenez@nd.edu

\begin{table}[h]
\begin{center}
\begin{tabular}{|c|c|c|c|c|}
\hline 
$D$ &$p_c$ &$\beta/\nu_\perp$& $\beta$ & $\nu_{||}/\nu_\perp$ \\
\hline\hline
0   & 0.077090(5) & 0.2523(3) & 0.2765 &1.577(4)   \\
\hline
0.1  & 0.10648(3)  & 0.503(6) & 0.546(6) & 2.04(4)    \\
0.5  & 0.12045(3)  & 0.430(2) & 0.468(2) & 1.86(2)    \\
0.85 & 0.13003(1)  & 0.412(2) & 0.454(2) & 1.77(2)    \\
\hline
BAW  &    -        & 0.497(5) & 0.922(5) & 1.74(1)    \\
\hline
\end{tabular}
\end{center}
\label{pcprat}
\noindent{Table I. Static exponents for the PCPD and the BAW model;
figures in parentheses denote uncertainties.
BAW results from Ref. \cite{zhong}.
}  
\end{table}

\begin{table}[h]
\begin{center}
\begin{tabular}{|c|c|c|c|}
\hline 
$D$ &$p_c$ &$\delta$& $\eta$ \\
\hline\hline
0.2 & 0.28526(1)  & 0.223(1) & 0.198(1)    \\
0.6 & 0.19324(6)  & 0.212(5) & 0.220(1)    \\
\hline
BAW  &    -       & 0.286(2)  & 0.286(2)    \\
\hline
\end{tabular}
\end{center}
\label{parallel}
\noindent{Table II. Dynamic exponents for the PCPD and BAW model.
BAW results from Ref. \cite{zhong}.
}  
\end{table}

\newpage
\noindent FIGURE CAPTIONS
\vspace{1em}

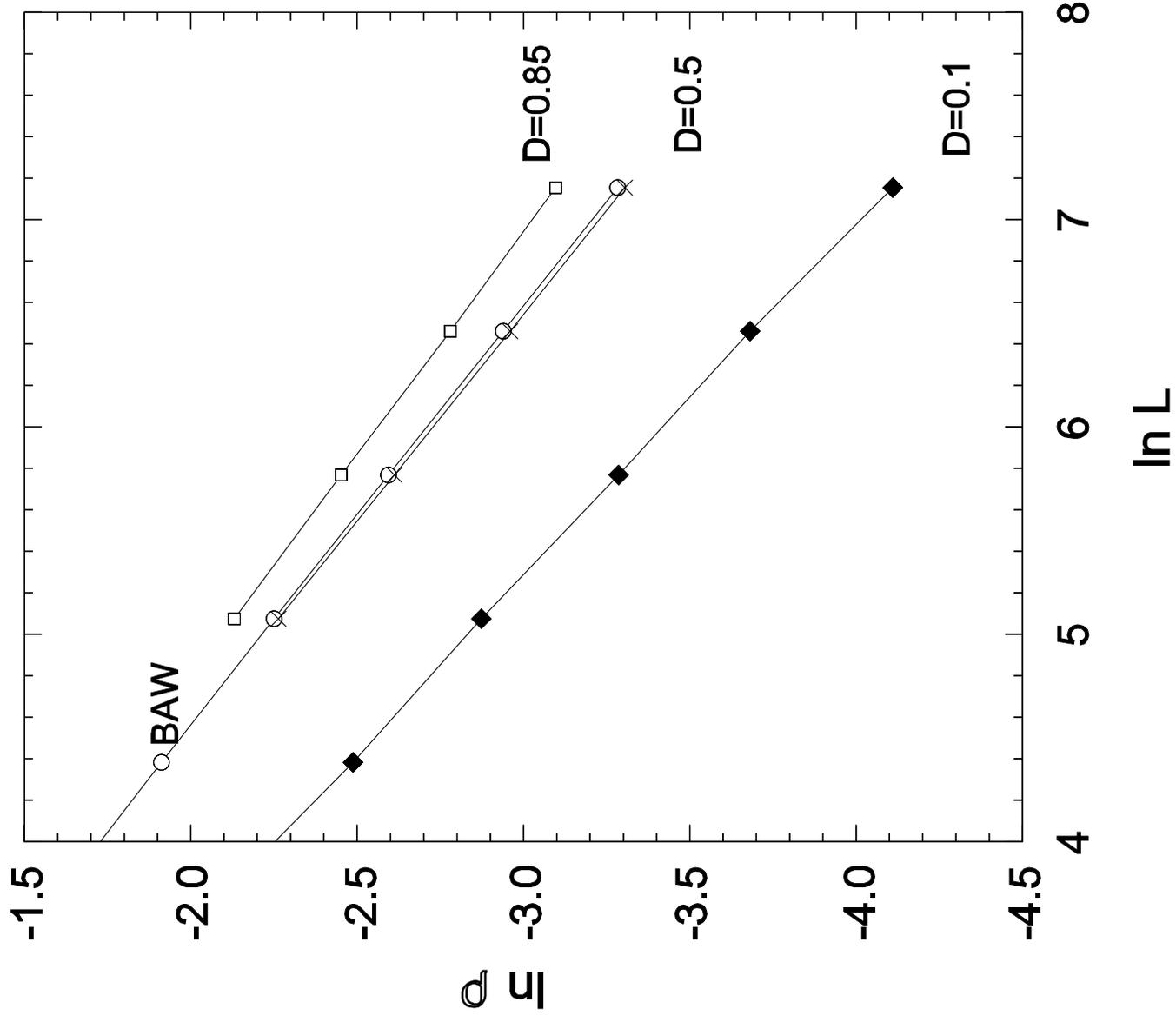
\noindent FIG. 1. Particle density $\rho$ versus system size at the critical point 
in the PCPD and the BAW model.
\vspace{1em}

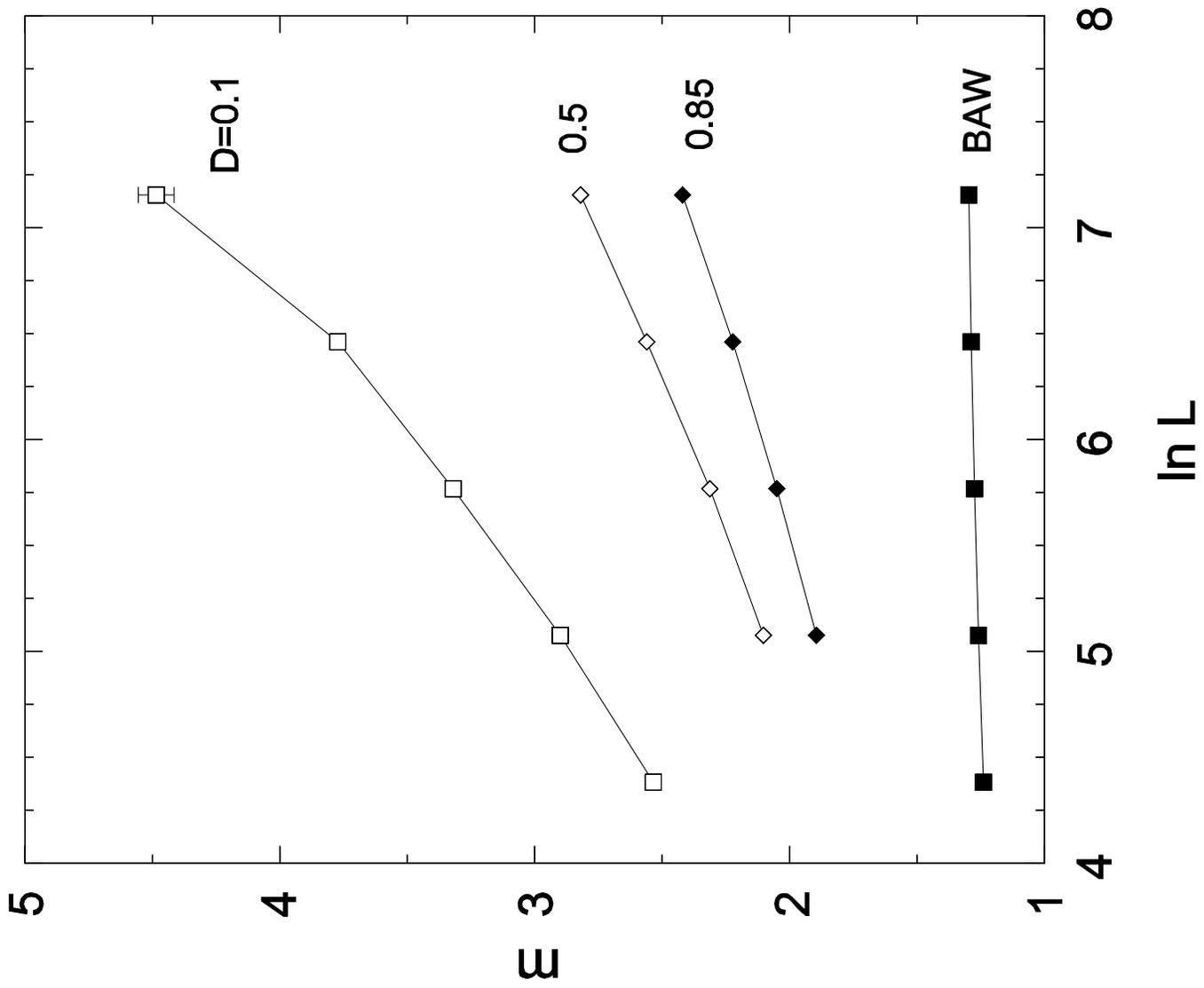
\noindent FIG. 2. Moment ratio $m$ versus system size at the critical point 
in the PCPD and the BAW model.
\vspace{1em}

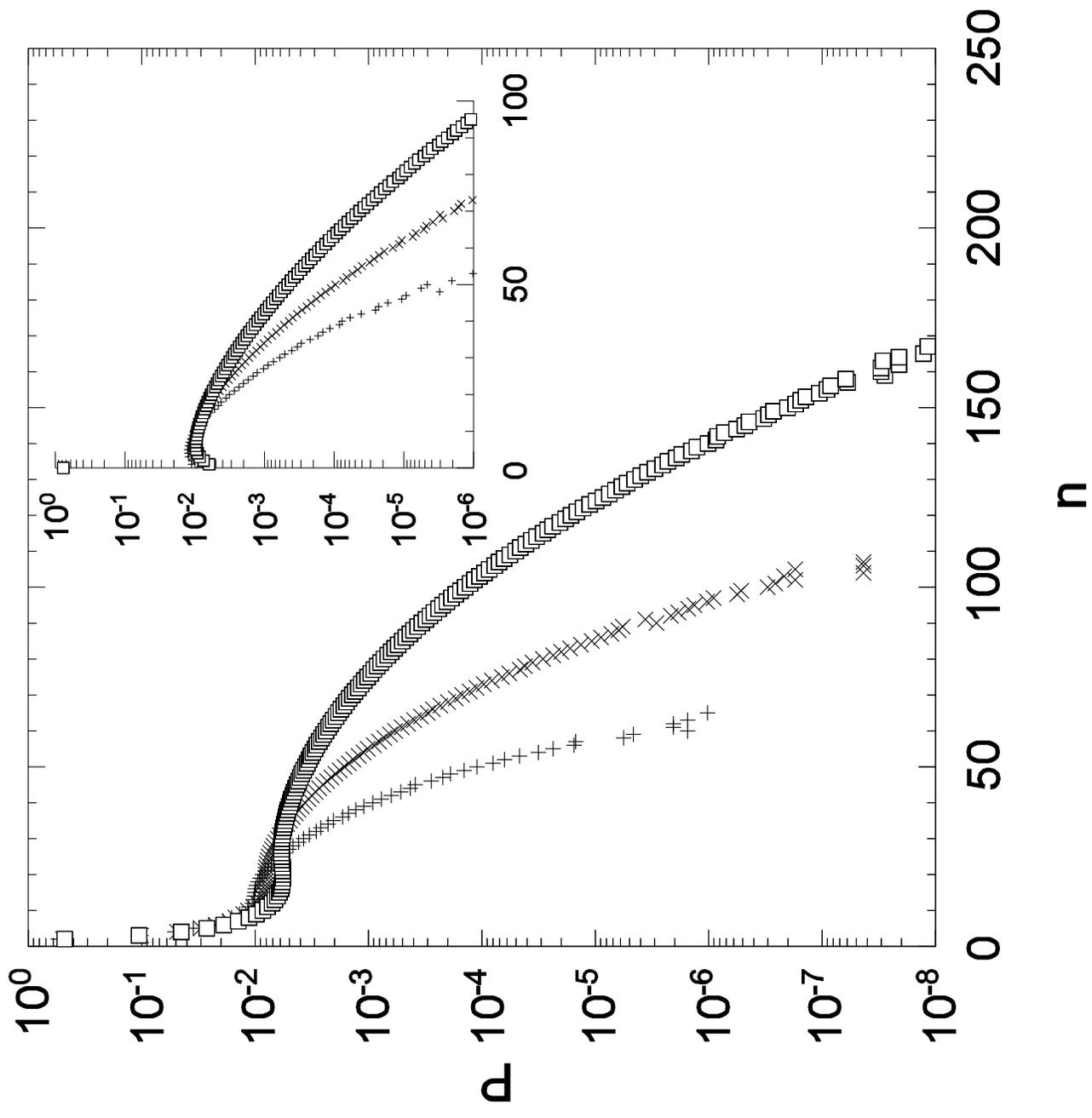
\noindent FIG. 3. Probability distribution of the number of particles $n$ for $D\!=\!0.1$.
+: $L\!=\!80$; $\times$: $L\!=\!160$; $\Box$: $L\!=\!320$.
The inset shows the corresponding probability distributions for the number of pairs, $n_p$.
Note that the most probable value of $n_p$ is zero.
\vspace{1em}

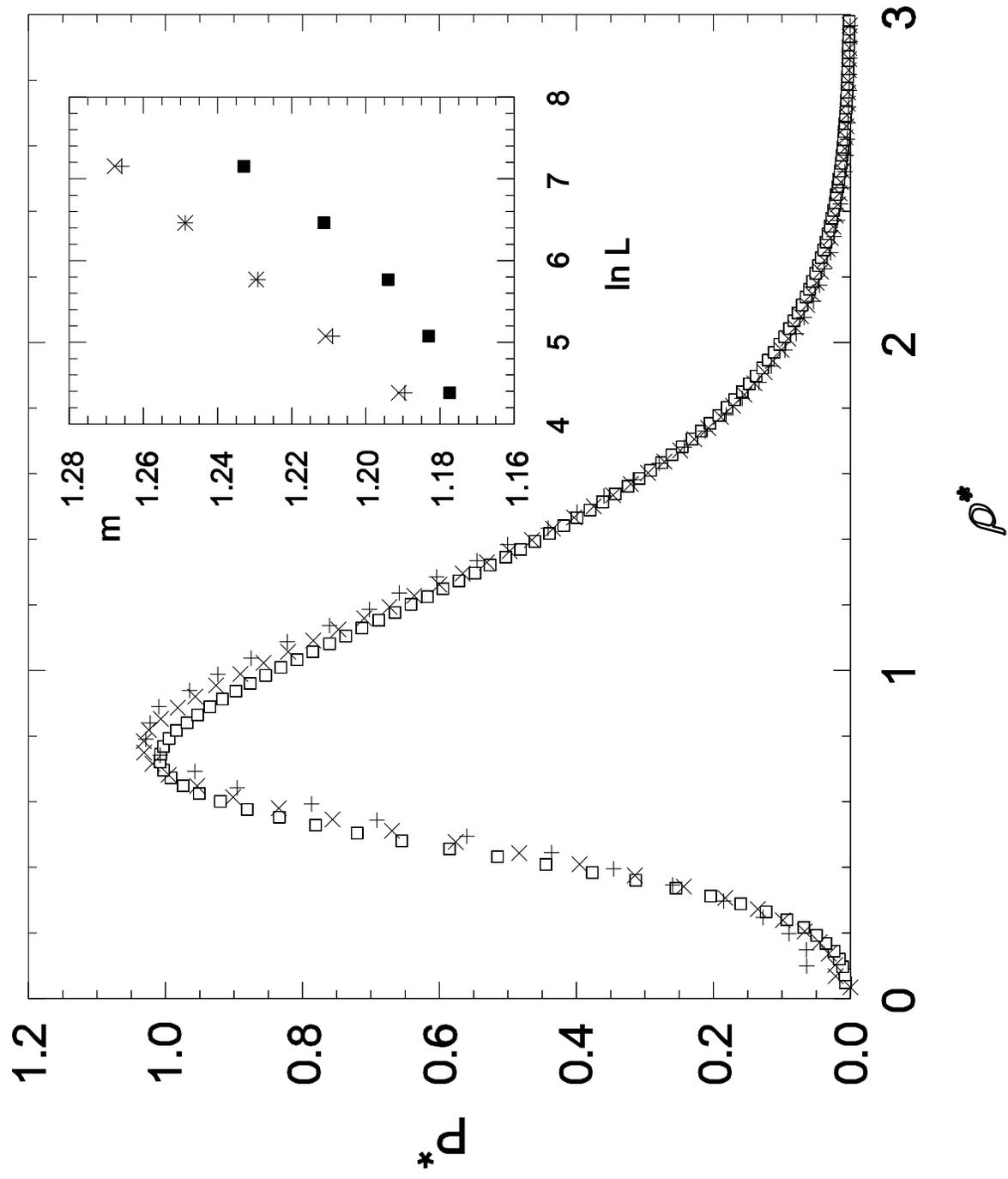
\noindent FIG. 4. Scaling plot of the probability distribution in the reactive sector for the same parameter
values as in Fig. 3.  Inset: moment ratio $m$ versus system size in the reactive sector;
filled squares: $D\!=\!0.1$; $+$: $D\!=\!0.5$; $\times$: $D\!=\!0.85$.
\vspace{1em}

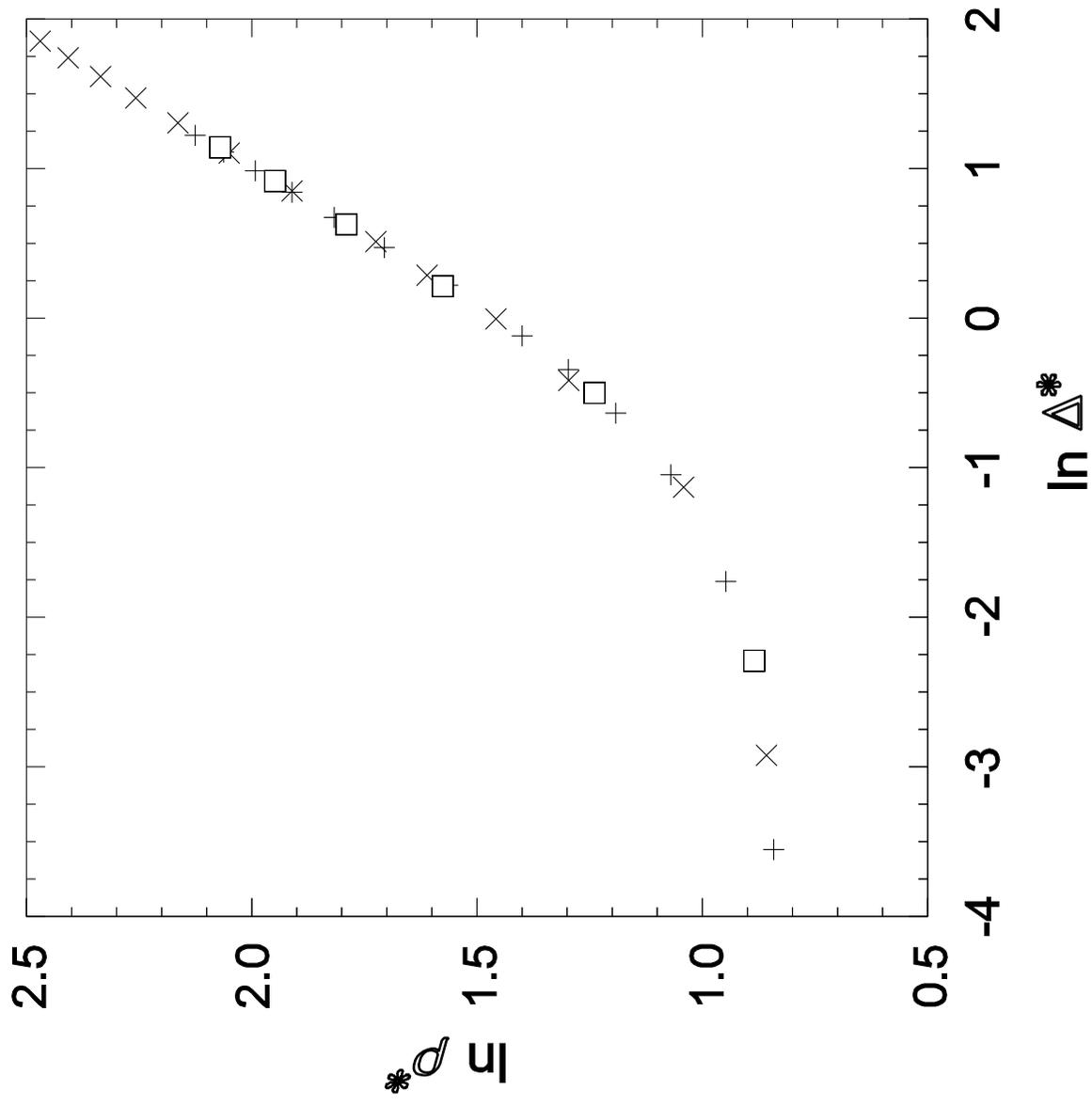
\noindent FIG. 5. Scaling plot of the order parameter in the reactive sector for $D\!=\!0.1$.
$+$: $L\!=\!640$; $\times$: $L\!=\!1280$; $\Box$: $L\!=\!2560$.

\end{document}